\documentclass[aps,prl,twocolumn,showpacs,groupaddress,floatfix]{revtex4}

\usepackage{graphicx,graphics}
\usepackage{dcolumn}
\usepackage{amsmath,amssymb,amsfonts}
\usepackage{latexsym,verbatim}
\usepackage{bm}
\usepackage{color}
\usepackage{ulem}
\usepackage[breaklinks=true,colorlinks,citecolor=blue,linkcolor=blue,urlcolor=blue]{hyperref}

\begin{document}

\title{Directional excitation of graphene surface plasmons}

\author{Fangli Liu}
\email{liuf0025@e.ntu.edu.sg}
\affiliation{School of Physical and Mathematical Sciences and Centre
  for Disruptive Photonic Technologies, Nanyang Technological
  University, 21 Nanyang Link, 637371, Singapore}

\author{Cheng Qian}
\affiliation{School of Computer Engineering, Nanyang Technological
  University, Nanyang Avenue, 639798, Singapore}

\author{Y.~D.~Chong}
\email{yidong@ntu.edu.sg}
\affiliation{School of Physical and Mathematical Sciences and Centre
  for Disruptive Photonic Technologies, Nanyang Technological
  University, 21 Nanyang Link, 637371, Singapore}

\begin{abstract}
We propose a scheme to directionally couple light into graphene
plasmons by placing a graphene sheet on a magneto-optical substrate.
When a magnetic field is applied parallel to the surface, the graphene
plasmon dispersion relation becomes asymmetric in the forward and
backward directions.  It is possible to achieve unidirectional
excitation of graphene plasmons with normally incident illumination by
applying a grating to the substrate.  The directionality can be
actively controlled by electrically gating the graphene, or by varying
the magnetic bias. This scheme may have applications in graphene-based
opto-electronics and sensing.
\end{abstract}


\maketitle

\noindent
{\it Introduction.---} Since the first exfoliation of graphene ten
years ago \cite{nov04}, this two-dimensional
material has attracted tremendous research interest, due to its unique
electronic \cite{rmp09}, mechanical \cite{mechan}, optical
\cite{optical} and thermal properties \cite{thermal}.  Its high
carrier mobility at room temperature  makes it a promising
material for post-silicon electronics, as well as for photonic and
opto-electronic devices.  Although an isolated layer of intrinsic
graphene interacts rather weakly with light (absorbing only $\sim
2.3\%$ of the incident intensity, independent of frequency
\cite{optical}), the interaction can be enhanced by external
cavity resonances \cite{engel_naturecommun_2012,furchi_nanolett_2012,
  fangli}, or by plasmonic resonances of the graphene surface itself
\cite{das, wang11, fengnian12, fengnian13}.  The latter approach,
which is the domain of the emerging field of ``graphene plasmonics'',
produces operating frequencies in the technologically-relevant
terahertz to mid-infrared range.  Compared to conventional metal
surface plasmons, graphene surface plasmons are subject to lower
propagation losses \cite{marin, kop,fei_nature_2012}.  They also have the virtue of being highly
tunable, via electrostatic or chemical doping of the graphene layer
\cite{wang11,fengnian12}.

This paper investigates the interesting possibility of
\textit{directionally} exciting graphene surface plasmons with the aid
of a magneto-optical substrate.  In order for graphene plasmons to be
excited by external illumination, surface modulations are typically
needed for wave-vector matching.  One approach is to pattern the
graphene into nano-ribbons, which localizes the plasmon resonances
\cite{wang11,fengnian12,fengnian13}.  Alternatively,
propagating graphene plasmons can be excited by patterning gratings
into the substrate \cite{aires}, or by applying elastic vibrations to
the graphene \cite{fahat,schie}. In principle, very high coupling
efficiencies (on the order of 50\%) can be achieved \cite{fahat}.
However, these methods excite plasmons in both directions along the
grating.  For numerous switching applications, it would be useful to
be able to control the propagation direction of the plasmons.  In the
literature on conventional surface plasmons, several methods for
directional excitation have recently been explored \cite{tam,
  dire,ting,xiao,lin,rod,huang}.  However, it is challenging to apply
most of these methods \cite{ting,xiao,lin,rod,huang} to graphene
plasmons, due to graphene's two-dimensional nature, as well as the
technical difficulty of patterning a meta-surface onto graphene
without harming the optical sheet conductivity by edge scattering
\cite{fengnian13}.

The system we investigate consists of a graphene layer on a
magneto-optical substrate magnetized parallel to the surface.  Unlike
other recent papers on magnetic graphene plasmons \cite{zhang06, cra,
  sad}, we do not magnetize the graphene sheet perpendicular to the
plane, which would induce Landau levels and an off-diagonal current
response.  As discussed below, such schemes do not lead to the desired
phenomenon of directional in-plane excitation.  In the present system,
the propagation of the graphene plasmon is made nonreciprocal by the
penetration of the plasmon mode into the magneto-optical substrate,
which is magnetically biased parallel to the surface and perpendicular
to the direction of propagation.  This configuration is reminiscent of
the work of Yu \textit{et al.}~\cite{shanhui}, who showed that a
surface plasmon at the interface between a dielectric and a magnetic
metal (likewise biased parallel to the surface) exhibits an asymmetric
dispersion relation.  Remarkably, because that system has a dispersion
relation with a frequency cutoff, it acts as a unidirectional
plasmonic waveguide over a finite frequency bandwidth below the
cutoff.  In the case of graphene plasmons, it is a well-known fact
that no such cutoff exists \cite{marin}.  Nonetheless, directionality
can be enforced at \textit{specific} frequencies by the application of
a grating, e.g.~with a modulated substrate surface.  Light incident at
an appropriately-chosen frequency can thus excite graphene plasmons in
one direction.  For a fixed operating frequency, it is even possible
to actively reverse the direction of excitation, either by switching
the magnetic field direction, or (more interestingly) by varying the
doping level of the graphene layer.  Such a device may have
applications in graphene-based opto-electronic and photonic devices,
such as optical absorbers, nano-sensors, and molecular detectors.

\begin{figure}
\centering \includegraphics[width=0.4\textwidth]{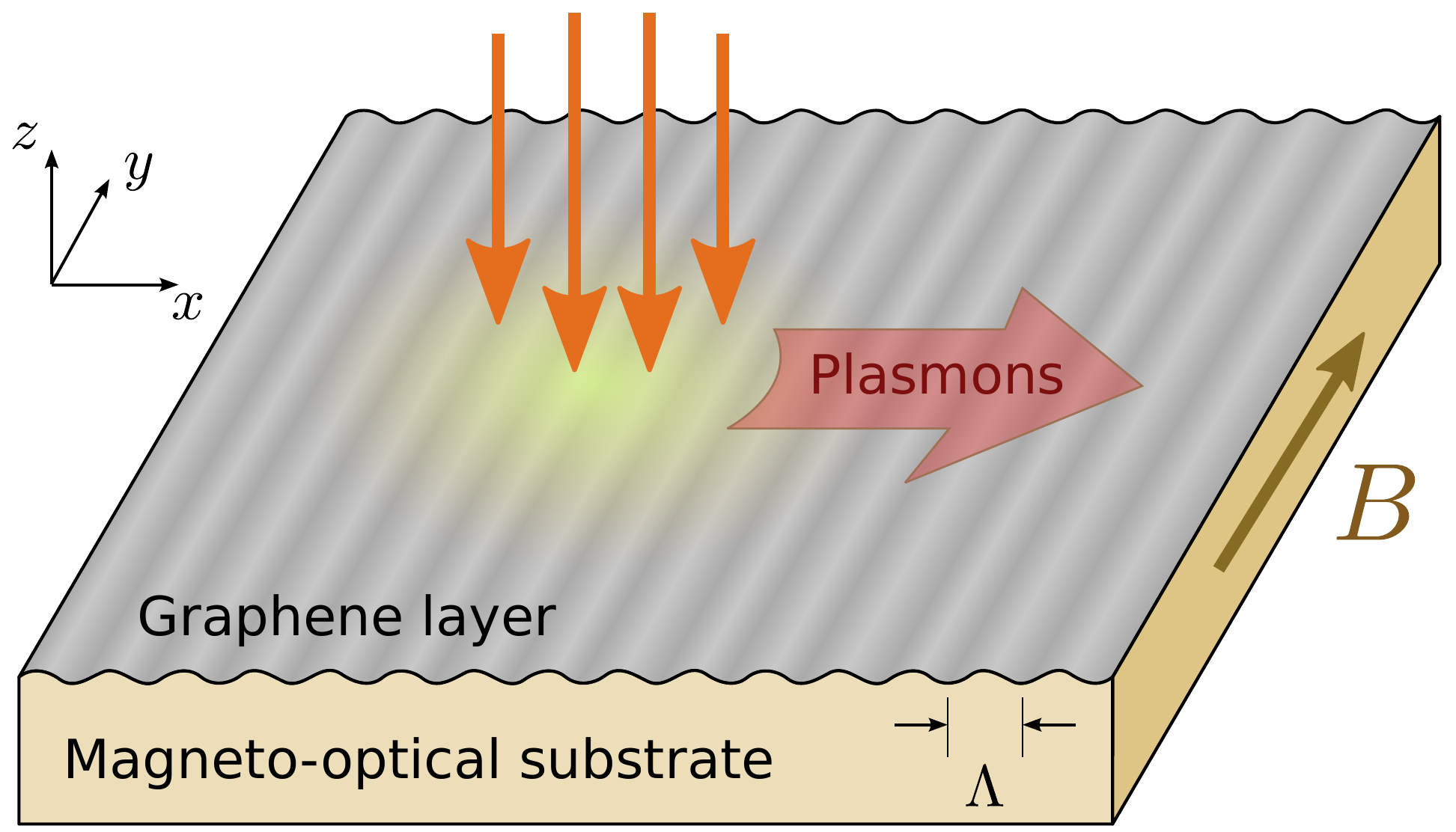}
\caption{Schematic of the device under investigation.  A graphene
  sheet lies on a grated magnetic-optical substrate which has period
  $\Lambda$.  A magnetic field is applied along the $\hat{y}$
  direction, parallel to the surface and perpendicular to the normal
  of the grating.  Normally incident monochromatic light excites
  surface plasmons propagating in a single direction. }\label{sche}
\end{figure}

{\it Asymmetric Dispersion Relation.---} Consider a graphene sheet
lying on an magnetic-optical substrate, as shown schematically in
Fig.~\ref{sche}.  A grating is etched onto the surface, with period
$\Lambda$.  If the modulation is sufficiently weak, it will not
significantly affect the dispersion relation of surface plasmon modes
\cite{anal}. Thus, we can derive the plasmon dispersion by taking the
limit of zero modulation (i.e., a flat surface).  When a magnetic
field applied in the $\hat{y}$ direction, the relative dielectric
constant of the substrate (medium 1) is
\begin{equation}
 \vec{\epsilon}=\begin{bmatrix}\epsilon_1 & \ 0& \ i\alpha \\ 0 &\epsilon_\parallel& 0 \\ -i\alpha & 0 & \epsilon_1
 \end{bmatrix}.
  \label{di1}
\end{equation}
Here, $\epsilon_1$ and $\epsilon_\parallel$ are the dielectric
components perpendicular to and parallel to the magnetization;
$\alpha$ is the magneto-optical component.  Note that the magnetic
field is applied in $y$ direction (parallel to the surface), so it
does not split the electronic states into Landau levels, unlike the
case where the magnetic field is applied through the graphene sheet
\cite{cra,zhang06, sad}.  Above the graphene, the dielectric constant
is $\epsilon_2$ (medium 2).  The graphene layer has isotropic sheet
conductivity $\sigma$.

\begin{figure}
\centering \includegraphics[width=0.45\textwidth]{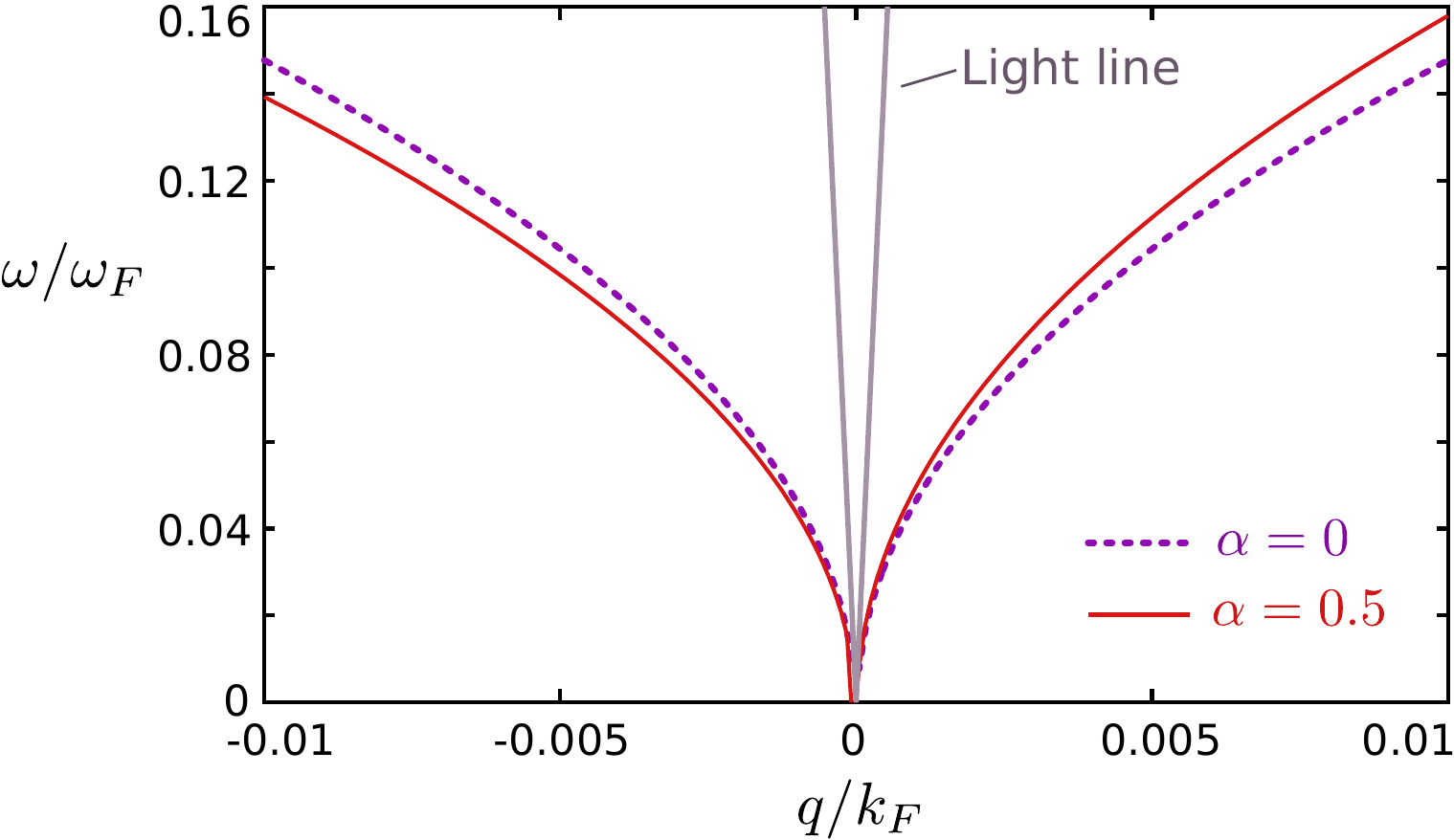}
\caption{Dispersion relations for graphene surface plasmons on
  magneto-optical substrates, calculated numerically from
  Eqs.~(\ref{disp})-(\ref{drude}).  The wavenumber and frequency are
  respectively scaled relative to $k_F$ and $\omega_F$, the Fermi
  wavenumber and frequency in the graphene sheet (which are both
  proportional to the Fermi level).  We assume a Fermi velocity of
  $10^6\mathrm{ms}^{-1}$ and $\tau = 0$ in the graphene sheet.  The
  relative dielectric constants $\epsilon_1 = 3$ and $\epsilon_2 = 1$,
  and the magneto-optical parameter $\alpha$, are taken to be
  frequency-independent.  For $\alpha = 0$, the dispersion relation is
  symmetric; setting $\alpha \ne 0$ makes it asymmetric. }\label{asym}
\end{figure}

We look for surface plasmon modes where the electric and magnetic
fields in medium 1 and 2 have the form
\begin{align}
\vec{E_j}(x,z) &=(E_{jx}, \ 0, \ E_{jz} ) e^{iqx} e^{-k_j \vert z\vert}\\
\vec{H_j}(x,z) &=(0, \ H_{jy}, \ 0 ) e^{iqx} e^{-k_j \vert z\vert},
\end{align}
where $q$ denotes the wavevector in the $\hat{x}$ direction, $k_j$ is
the decay constant in the $z$ direction, for medium index $j=1,2$.  We
plug these expressions into Maxwell's equations, assume all the
time-dependent fields are harmonic with frequency $\omega$, and impose
the boundary conditions $E_{1x}= E_{2x}$ and $B_{1y}=B_{2y} -\sigma
E_{1x}$ along the graphene layer.  The resulting dispersion relation
is:
\begin{equation}
\frac{\epsilon_2}{k_2}+ \frac{\epsilon_1^2-\alpha^2}{k_1 \epsilon_1+
  q\alpha} + \frac{i\sigma}{\omega \epsilon_0}=0,
\label{disp}
\end{equation}
where $\epsilon_0$ is the permittivity of free space and
\begin{align}
k_1&=\sqrt{q^2-\left(\frac{\epsilon_1^2-\alpha^2}{\epsilon_1}\right) \,\frac{\omega^2}{c^2}} \\
k_2&= \sqrt{q^2-\epsilon_2\,\frac{\omega^2}{c^2}}.
\label{k2}
\end{align}
Evidently, $k_1$ and $k_2$ only depend on the absolute value of $q$,
and do not depend on the direction of propagation.  However, the
second term of Eq.~(\ref{disp}) has a denominator which involves $q$
rather than $q^2$.  This is the source of the magneto-optically
induced asymmetry.

Graphene's sheet conductivity can be modeled by \cite{primer}
\begin{equation}
  \sigma(\omega)= \frac{e^2 E_F}{\pi \hbar^2}\, \frac{i}{\omega+i/\tau},
  \label{drude}
\end{equation}
where $E_F$, $\tau$, $\hbar$, and $e$ are the doping level, damping
time, Planck constant, and electron charge respectively.
Fig.~\ref{asym} shows the asymmetric dispersion relation computed
numerically from Eqs.~(\ref{disp})-(\ref{drude}), assuming
frequency-independent dielectric components.

We will work in the ``non-retarded'' regime, where the mode lies well
below the light line and is well-confined \cite{marin}.  For $1/\tau
\ll \omega \ll qc$, Eqs.~(\ref{disp})-(\ref{drude}) simplify to
\begin{equation}
\omega\approx\sqrt{\frac{e^2 E_F }{\pi \hbar^2 \epsilon_0}\cdot
  \frac{\mid q\mid}{\epsilon_1+\epsilon_2 \mp \alpha}},
\label{adisp}
\end{equation}
where $\pm$ denote right and left propagation respectively.  For
$\alpha=0$, Eq.~(\ref{adisp}) reduces to the usual square-root
dispersion relation for graphene surface plasmons \cite{marin}.  For
$\alpha \ne 0$, the dispersion relation is asymmetric; for fixed
$|q|$, there are two different values for $\omega$.  Note that
$\alpha$, $\epsilon_1$, and $\epsilon_2$ can also depend implicitly on
$\omega$.

From Eq.~(\ref{adisp}), we obtain the frequency difference ratio
\begin{equation}
  \frac{\delta\omega}{\omega_0}\sim\frac{\alpha}{\epsilon_1+\epsilon_2},
  \label{difference ratio}
\end{equation}
where, for fixed $|q|$, $\delta\omega$ is the frequency difference
between right- and left-moving plasmons.  This ratio depends only on
the relative strength of the magnetic-optical component in the bulk
media.  It does not depend on the graphene sheet conductivity, which
affects only $\omega_0$, the unmagnetized graphene surface plasmon
frequency.

For a typical magneto-optical material, the dielectric components
defined in (\ref{di1}) can be modeled by \cite{qijie1,qijie2,riv}
\begin{align}
  \epsilon_1(\omega) &= \left(1-\frac{\omega_p^2}{\omega^2-\omega_c^2}\right)
  \epsilon_\infty  \label{epsn1}
  \\ \alpha(\omega) &= \frac{\omega_p^2 \omega_c }{\omega
    (\omega^2-\omega_c^2)} \; \epsilon_\infty, \label{alpha}
\end{align}
where $\epsilon_\infty$ is the limiting permittivity at high
frequencies, $\omega_p$ is the bulk plasma frequency, and $\omega_c$
is the cyclotron frequency induced by the magnetic bias.  Assuming
$\epsilon_2 = 1$ and weak magnetization ($\omega_c \ll
\omega,\omega_p$), combining Eqs.~(\ref{difference
  ratio})-(\ref{alpha}) gives
\begin{equation}
  \delta \omega \sim \omega_c \cdot
  \frac{\omega_p^2/\omega_0^2}{\epsilon_\infty^{-1} + 1 -\omega_p^2/\omega_0^2}.
  \label{fdr}
\end{equation}

It is useful to compare these results to the ``magnetic surface
plasmon'' which occurs at the interface between a dielectric and a
magnetic metal \cite{shanhui}.  That situation corresponds to setting
$\sigma = 0$ in Eq.~(\ref{disp}); in order for a confined state to
exist in the absence of the conducting layer, the magneto-optical
material must then be metallic ($\epsilon_1 < 0$).  In the large-$q$
regime, the magneto-optical frequency difference is found to be
$\delta\omega \sim \omega_c$, with no leading-order dependence on
$\epsilon_\infty$ or $\omega_p$.  Because the magnetic surface
plasmon's dispersion relation has a frequency cutoff, propagation
becomes unidirectional in a finite frequency range near the cutoff
\cite{shanhui}.  In the present case, however, the confinement is
provided by the graphene layer, and we can take the magneto-optical
material to be non-metallic ($\epsilon_1 > 0$); because the dispersion
relation has no cutoff, directionality has to be imposed by other
means, such as a grating.

\begin{figure}
\centering \includegraphics[width=0.45\textwidth]{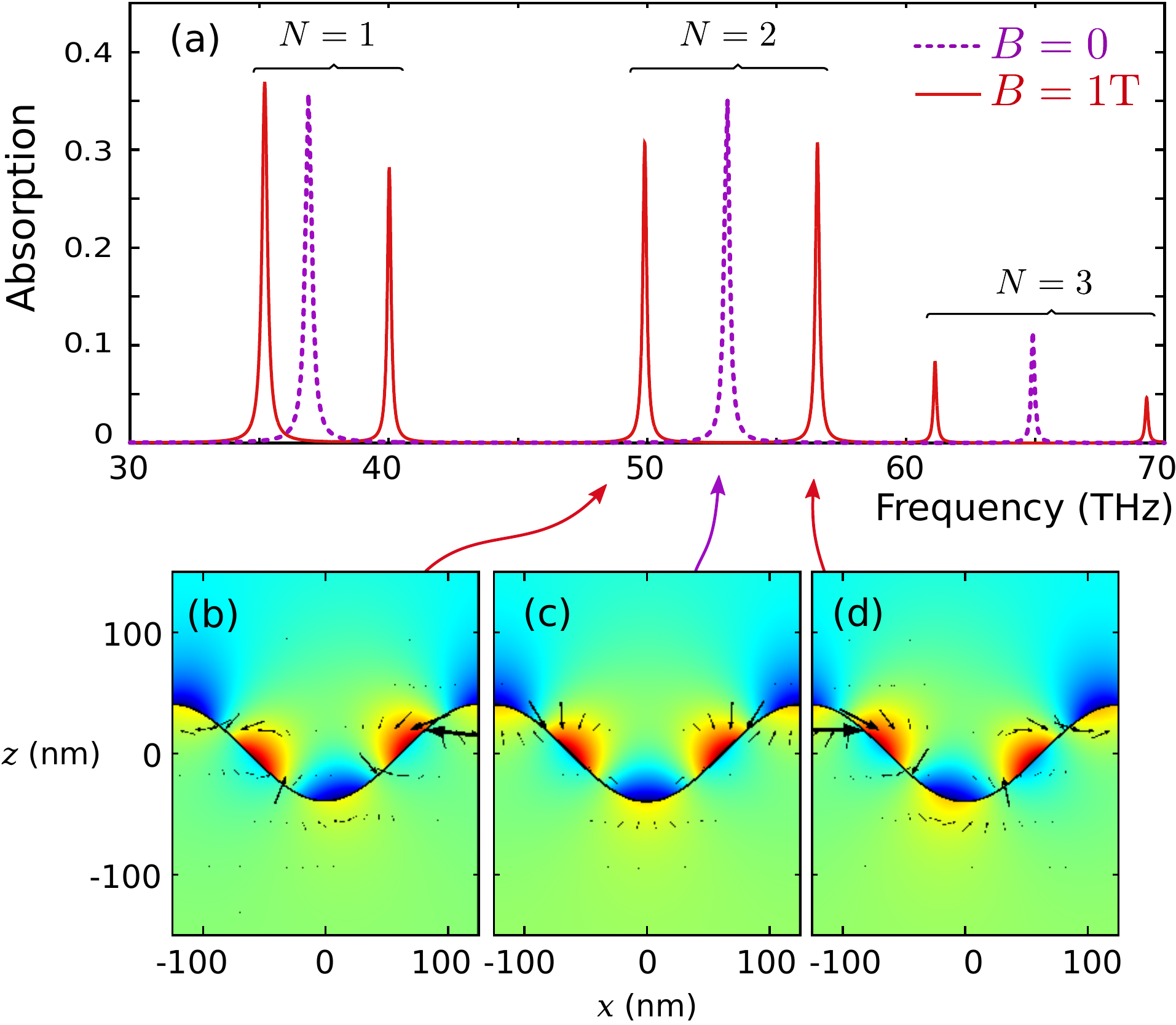}
\caption{(a) Absorption spectrum of graphene surface plasmon
  resonances, obtained by full-wave simulations of the system shown in
  Fig.~\ref{sche}.  With zero applied magnetic field (blue triangles),
  there is a single absorption peak, corresponding to bi-directional
  excitation of surface plasmons.  With a 1T applied magnetic field
  (red circles), there are two distinct absorption peaks,
  corresponding to plasmons propagating with wave-vector $\pm
  2\pi/\Lambda$, where $\Lambda$ is the grating period.  (b)-(d) Plots
  of $H_z$ at each of the resonances, with black arrows indicating the
  local Poynting vector. }\label{simu}
\end{figure}


{\it Directional coupling.---} To excite the graphene surface
plasmons, we consider applying a weak periodic modulation to the
surface of the substrate.  The grating couples incident light with
transverse wavevector component $k_0$ to surface plasmons with
wave-vector $k_{\textrm{spp}}= k_0 \pm 2\pi N/\Lambda$, where
$\Lambda$ is the grating period and $N \in \mathbb{Z}^+$.  In the
non-retarded regime, the directionality of the incident light has
little effect on the coupling since $k_0 \ll k_{\textrm{spp}}$ (as
indicated in Fig.~\ref{asym}).  Henceforth, we assume normal incidence
($k_0 = 0$) for simplicity.

For nonzero magnetic bias, the incident light couples to left- and
right-moving surface plasmons at different resonance frequencies.  To
demonstrate this effect, we perform full-wave finite element
simulations of Maxwell's equations (using the Comsol Multiphysics
software package) for the setup of Fig.~\ref{sche}.  The results are
shown in Fig.~\ref{simu}.  For the substrate, we use indium antimonide
(InSb), which has been shown to have large magnetic-optical response
at terahertz and mid-infrared frequencies \cite{qijie1,qijie2,riv}. Its dielectric tensor components can be modeled by
Eqs.~(\ref{epsn1})-(\ref{alpha}), with $\epsilon_\infty=15.68$ and
$\omega_c= eB/m^*$ (where $m^*=0.014m_e$ is the effective mass).  The
bulk plasma frequency is given by $\omega_p^2= Ne^2/(\epsilon_{\infty}
m^* \epsilon_0)$, where $N=5.5\times10^6\,\mu\mathrm{m}^{-3}$.  For an
operating frequency of 50 THz and magnetic bias of $B=1$T, this
results in $\epsilon_1 \approx 3$ and $\alpha \approx 0.5$.  The
grating has period $\Lambda = 250$ nm and modulation amplitude $40$ nm
(note that these length scales are sufficiently large that quantum
effects can be safely ignored \cite{nonlocal}).  To model the
graphene sheet, we use a thin layer of thickness $d \approx 0.3$ nm
with effective dielectric constant $\epsilon_{\mathrm{eff}} = 1 + i
\sigma/(\epsilon_0 \omega d)$.  The graphene conductivity parameters
are taken to be $E_F=1$ eV and $\tau=10^{-12}\mathrm{s}^{-1}$, in line
with typical experimental values for doped graphene on a semiconductor
substrate.

In Fig.~\ref{simu}(a), we see that in the absence of an applied
magnetic field is applied, there are absorption peaks near 37 THz, 53
THz, 65 THz \dots, corresponding to the excitation of surface plasmons
for different values of $N$.  The field distribution of $H_y$ for the
$N = 2$ mode is plotted in Fig.~\ref{simu}(c), showing that it is a
tightly-confined surface plasmon mode.  The Poynting vectors,
indicated by black arrows in this plot, reveal that the mode is
symmetric (i.e.,~zero net energy flow along the $\hat{x}$ axis).  When
a magnetic field is applied, each absorption peak splits into two, one
below the original frequency and one above, with absorption
coefficients comparable to the original peak.  The field distributions
for the $N = 2$ modes are plotted in Fig.~\ref{simu}(b) and (d), and
the Poynting vectors indicate the modes are directional, transporting
energy in the $-\hat{x}$ and $+\hat{x}$ directions respectively.  When
the direction of the applied magnetic field is reversed, $\alpha$
switches sign.  Thus, at a fixed operating frequency, we can actively
control the directionality of the excited surface plasmons.

\begin{figure}
\centering \includegraphics[width=0.43\textwidth]{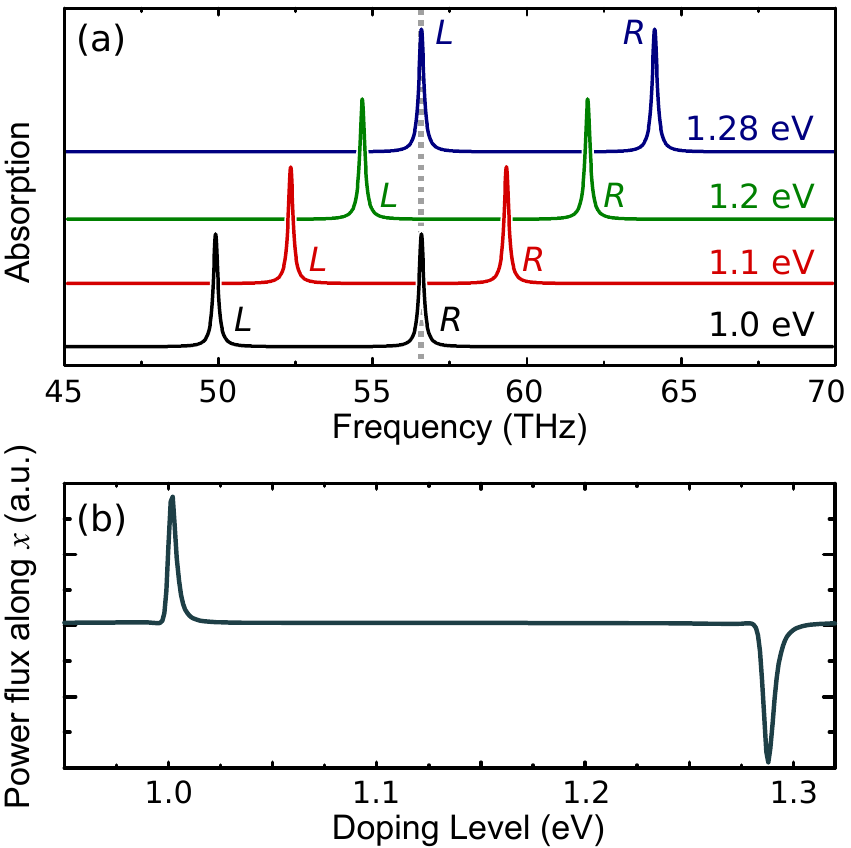}
\caption{ Active control of plasmon excitation directionality. (a) The
  resonance peaks, which correspond to left-moving ($L$) and
  right-moving ($R$) plasmons, can be shifted by tuning the doping
  level of the graphene layer.  Hence, at a fixed operating frequency
  (vertical dashes), we can excite right-moving plasmons at a 1 eV
  doping level, or right-moving plasmons at 1.28 eV. (b) The power
  flux through one unit cell in the $x$ direction (calculated by
  integrating the Poynting vector) versus the doping level, for a
  fixed excitation frequency of 56.6 THz.  This confirms that the
  plasmons propagate to the left and right at 1 eV and 1.28 eV doping
  levels respectively. These results were obtained from finite-element
  simulations with all other parameters the same as in
  Fig.~\ref{simu}. }\label{dire}
\end{figure}

Another interesting way to control the directionality of the excited
surface plasmons is to exploit the extraordinary tunability of
graphene's electrical properties.  The value of $\omega_F$ in
graphene, which enters into the dispersion relation of the surface
plasmon via Eq.~(\ref{adisp}), can be controlled very precisely by
electrical doping (or a combination of electrical and chemical
doping).  In Fig.~\ref{dire}, we give an example where, at a single
operating frequency ($56.6 Thz$), the directionality of the excited
surface plasmons can be reversed by switching the doping level from $1
eV$ to $1.28 eV$.  This tunability may be useful for applications in
switchable plasmonics.

{\it Discussion.---} The setup we have considered in this paper is
different from the one considered in most previous works on magnetic
plasmons in graphene, where the magnetic field is applied
perpendicular to the plane of the graphene sheet.  In that latter
setup, the graphene bandstructure forms Landau levels \cite{zhang06,
  cra, sad}, resulting in a complicated splitting of the plasmon
frequencies \cite{hugen}.  However, this splitting only affects the
transmission coefficient of circular polarized incident light; it does
not excite the plasmon in a specific \textit{in-plane} direction,
which is the phenomenon we are interested in.  Recently, Xiao {\it
  et~al}~have studied a setup where the magnetic field is applied
in-plane \cite{linxiao}.  However, they treated the case of a stack of
graphene sheets several microns thick, and considered the
magneto-optical effect occurring in that stack, in a manner directly
analogous to the magnetic surface plasmons of Ref.~\cite{shanhui}.
Such a scheme is not practically realizable or relevant to the case of
monolayer graphene.  In our work, the magneto-optical effect is due to
the substrate material, not the graphene layer.

In our simulations, we have chosen to focus on the simplest mechanism
for coupling incident light to the surface plasmons, i.e.~by
modulating the substrate surface.  The properties of graphene have
previously been shown to be strongly affected by the choice of
substrate \cite{dean, shaffique, ong}, and the feasibility of
patterning graphene on a magneto-optical substrate, such as InSb, is
an open experimental question.  We emphasize, however, that details
about the implementation of the grating mainly affect the coupling
efficiency, and largely do not alter our analysis of the dispersion
relation of magneto-optical graphene surface plasmons.  For instance,
one might instead deposit a dielectric grating on top of the graphene,
or use graphene ribbons instead of a contiguous graphene sheet
\cite{wang11}.

In conclusion, we have presented a simple scheme for directionally
exciting graphene surface plasmons, based on the asymmetry in the
surface plasmon dispersion relation due to a magneto-optical
substrate.  The directionality can be actively controlled, such as by
electrically gating the graphene layer.  This proposal may be useful
as the coupling mechanism for more complex graphene plasmonic devices.
It would also be interesting to explore the possibility of using an
alternative conducting material, such as a metallic thin film, in
place of the graphene; the surface plasmons which arise in such
dielectric/thin metal/magneto-optical dielectric structures may have
useful applications outside the context of graphene-based
technologies.

{\it Acknowledgements.---}We thank X.~Lin and Q.~J.~Wang for helpful
discussions.  This research was supported by the Singapore National
Research Foundation under grant No.~NRFF2012-02, and by the Singapore
MOE Academic Research Fund Tier 3 grant MOE2011-T3-1-005.

\end{document}